\documentclass[journal,11pt,onecolumn,draftclsnofoot,]{IEEEtran}

\usepackage{tikz}
\usetikzlibrary{shapes,arrows}
\usepackage{caption}
\usepackage{amsmath}
\usepackage{float}
\usepackage{mhchem}
\usepackage{cleveref}

\begin{document}
	%
	\title{A Fluid Dynamics Approach to Channel Modeling in Macroscale Molecular Communication}
	%
	%
	%
	
	\author{Fatih~Gulec, Baris~Atakan\thanks{This work was supported by the Scientific and Technological Research Council of Turkey (TUBITAK) under Grant 119E041.}
		\thanks{The authors are with the Department
			of Electrical and Electronics Engineering, Izmir Institute of Technology, 35430, Urla, Izmir, Turkey. (email: fatihgulec@iyte.edu.tr; barisatakan@iyte.edu.tr)}}
	\maketitle
	
	\begin{abstract}
		In this paper, a novel fluid dynamics-based approach  to channel modeling, which considers liquid droplets as information carriers instead of molecules in the molecular communication (MC) channel, is proposed for practical macroscale MC systems. This approach considers a two-phase flow which is generated by the interaction of droplets in liquid phase with air molecules in gas phase. Two-phase flow is combined with the signal reconstruction (SR) of the receiver (RX) to propose a channel model. The SR part of the model quantifies how the accuracy of the sensed molecular signal in its reception volume depends on the sensitivity response of the RX and the adhesion/detachment process of droplets. The proposed channel model is validated by employing experimental data.
	\end{abstract}
	\IEEEpeerreviewmaketitle
	
	\section{Introduction}
	\IEEEPARstart{M}{olecular} communication (MC) is a prominent communication paradigm  which can pave the way for potential macroscale applications, e.g., monitoring of a threatening molecular source. The experimental pioneering study given in \cite{farsad2013tabletop} shows that a MC link between an alcohol sprayer as the transmitter (TX) and an alcohol sensor as the receiver (RX) can be accomplished. This study is improved via multiple input multiple output (MIMO) technique in \cite{koo2016molecular}. \cite{farsad2017novel} proposes a platform using the pH level of chemicals to encode information symbols. In  \cite{unterweger2018experimental}, a platform which employs magnetic nanoparticles as information carrier molecules is proposed. \cite{ mcguiness2018parameter} introduces an odor generator and a mass spectrometer as the TX and RX for macroscale MC, respectively.
	
	In a practical macroscale MC application, employing an accurate channel model enables a more efficient information transfer between the TX and RX. In addition, channel parameters such as the distance between the TX and RX can be estimated via accurate channel models. As an exemplary application, an infected human emitting droplets into the air through sneezing or coughing can be considered as a molecular TX in public places \cite{khalid2019communication}. By deploying biological sensors as  molecular RXs, practical macroscale channel models can be employed to estimate the location of infected people. In \cite{farsad2014channel} and \cite{kim2015universal}, channel models for the platform in \cite{farsad2013tabletop} are derived by modifying the solution of the diffusion equation and model coefficients are estimated by fitting experimental data. Since the models in \cite{farsad2014channel} and \cite{kim2015universal} are based on the molecule diffusion assumption, correction factors are needed to modify the diffusion equation.  In these studies, it is not clearly known what fitted parameters of the modified diffusion equation are physically related to. Since the TX sprays liquid droplets rather than  molecules and droplets are larger with respect to molecules, only diffusion is not sufficient for channel modeling in practical scenarios. Thus, initial velocities and interactions of droplets with  air molecules and the surface of the sensor (RX) need to be considered. Hence, a fluid dynamics perspective is required for a more accurate channel model. 
	
	In contrast to the models in \cite{farsad2014channel} and \cite{kim2015universal} based on the diffusion of molecules, this paper proposes a more realistic system model based on a  fluid dynamics approach and reveals the physical meanings of the channel parameters. In this approach, droplets sprayed by the TX are considered as information carriers in the channel rather than diffusing molecules. The interaction of these droplets with air generates a two-phase flow where the first phase is liquid and the second phase is gas. In two-phase flow, droplets and air molecules move together in liquid and gas phases, respectively. The TX is modeled as a directed emitter with a predefined beamwidth. Droplets are assumed to move in a cone shaped volume determined by this beamwidth. The model also quantifies how accurate the transmitted signal is reconstructed by the RX by examining movements of  droplets, interactions among droplets and surface of the sensor. Therefore, this part of the model is called as signal reconstruction (SR). The SR involves two consecutive processes which are the adhesion/detachment process of droplets and the sensitivity response of the sensor.  The resulting end-to-end system response is derived and validated by experimental data. It is revealed that the end-to-end system response of a practical MC system depends on the distance,  parameters of the sensor measurement circuit, sensitivity characteristic of the sensor, beamwidth of the TX, spray coefficient and reaction rate constants in the adhesion/detachment process. 
	\section{A Practical Channel Model Based on Fluid Dynamics and Signal Reconstruction} \label{Model}
	In this section, we explain our experimental setup and introduce the fluid dynamics-based end-to-end system model. 
	\subsection{Experimental Setup}
	\label{Exp_Setup} 
\begin{figure}[h]
		\centering
		\scalebox{0.85}{\includegraphics{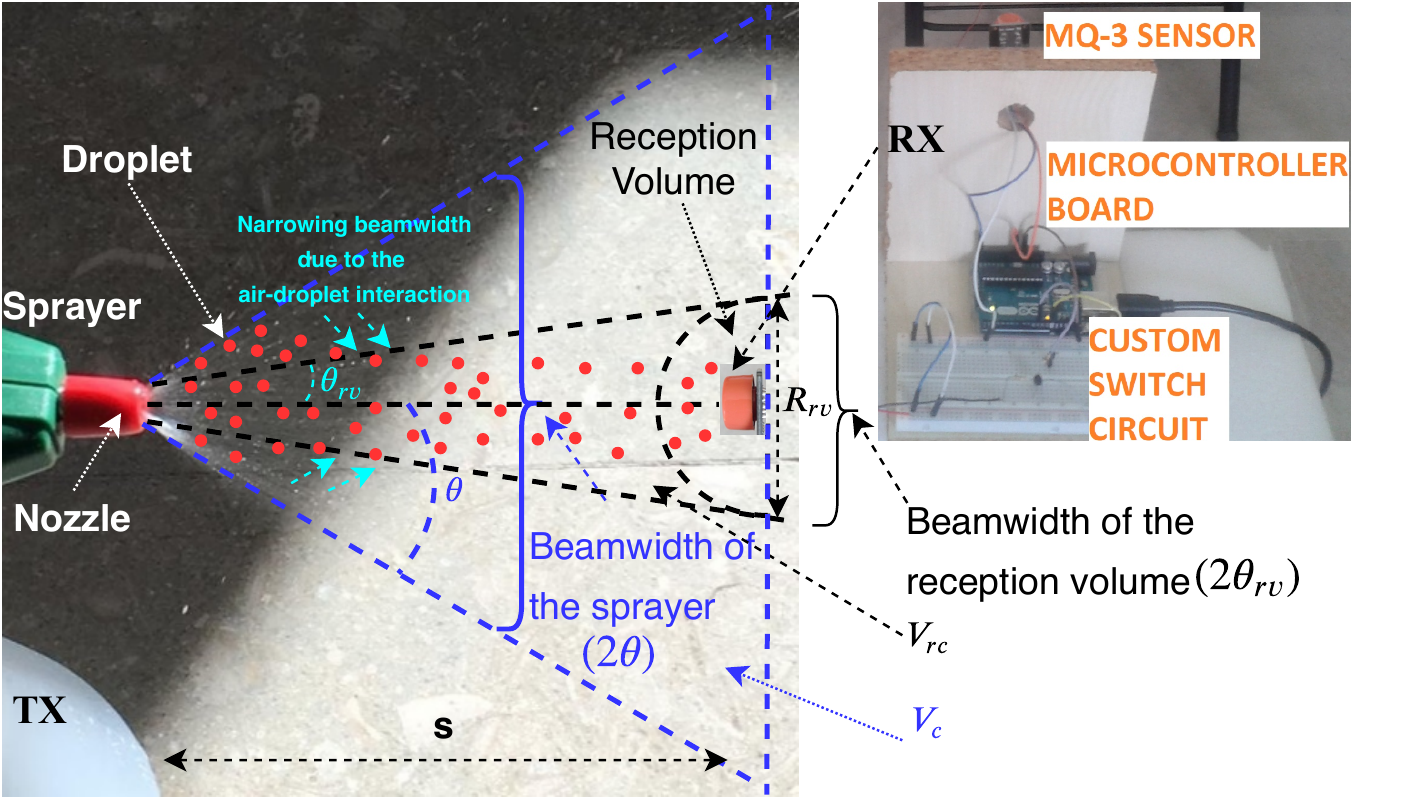}}
		\caption{The experimental setup and parameters.}
		\label{RV}
	\end{figure}

	In the experimental setup given in Fig. \ref{RV}, the TX is an electric sprayer which emits ethanol droplets and the RX is an MQ-3 alcohol sensor. The TX and RX are controlled via an Arduino Uno microcontroller board by the help of a  switch circuit. The TX sprays droplets with an initial velocity during the emission time ($T_e$). In contrast to experimental setups in \cite{farsad2014channel} and \cite{kim2015universal}, no fan is employed to drift droplets towards the RX. The TX and RX are aligned on the horizontal axis.
	\subsection{End-to-End System Model}
	\label{Channel}
\begin{figure}[h]
		\centering
		\scalebox{0.55}{\includegraphics{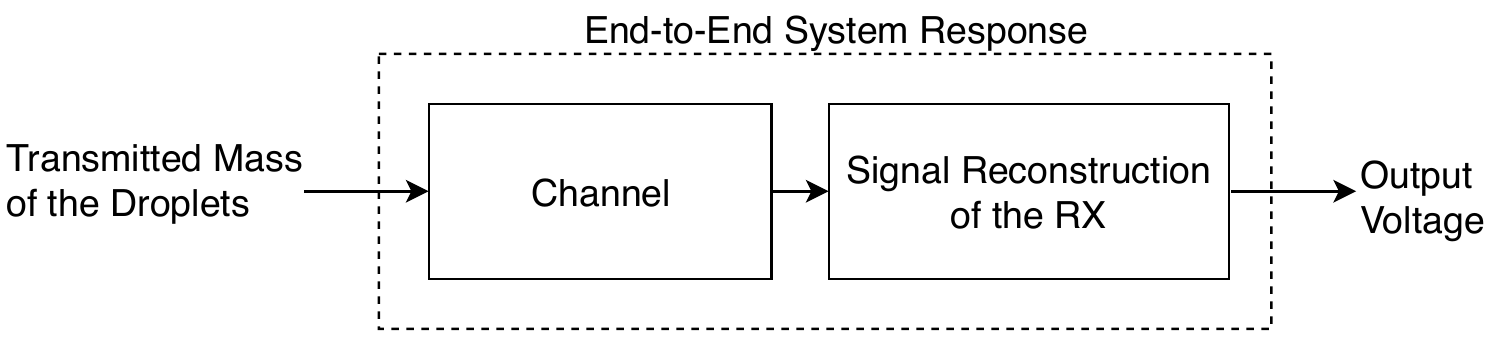}}
		\caption{Block diagram of the end-to-end system model.}
		\label{End}
	\end{figure}

	For the system model shown in Fig. \ref{End}, the transmitted mass of droplets is taken as the input and the measured sensor voltage is the output of the system. The end-to-end system impulse response ($E_{out}(t)$) is defined as the system's output to a short spray emission along $T_e$ (without a fan)  which can be considered as an impulsive input signal. For the derivation of $E_{out}(t)$, the propagation time of droplets ($t_0$) is needed to be estimated. However, we only focus on the signal after $t_0$. Solely, the effect of droplet propagation on the signal in the reception volume (RV) of the RX is investigated. In addition, the initial offset voltage of the sensor is eliminated by subtracting the sensor voltage at $t_0$ from the whole signal. 
	\subsubsection{The Effect of Two-Phase Flow on the Initial Concentration in the Reception Volume} 
	As shown in Fig. \ref{RV}, droplets are assumed to move in a cone-shaped volume \cite{ghosh1994induced} and the beamwidth ($2\theta$) is defined as the initial spraying angle of the sprayer's nozzle. The interaction among droplets and air molecules creates a two-phase flow where droplets and air molecules move together as a mixture \cite{ishii2010thermo}. Here, the first phase is the liquid phase of droplets and the second phase is the gas phase of air molecules. Due to this interaction, the majority of droplets move in a narrower beamwidth, as they propagate in the channel \cite{ghosh1994induced}. This narrower beamwidth, denoted by $2\theta_{rv}$, is assumed to encompass the RV as illustrated in Fig. \ref{RV}. Hence, the beam of droplets forms two concentric cones. The volume and base diameter of the inner cone are denoted by $V_{rc}$ and $R_{rv}$, respectively. The volume of the outer cone is $V_c$ and the distance between the TX and RX is $s$ as depicted in Fig. \ref{RV}. 
	
	When the spatial distribution of droplets is assumed to be homogeneous, the mass of droplets in the inner cone can be found via multiplying the total transmitted mass, i.e., $m_{TX}$, by the ratio of the cone volumes, i.e., $V_{rc}/V_c$. However, due to the interactions among droplets and air molecules, the propagation of droplets is far from homogeneity \cite{ghosh1994induced}. Therefore, in order to quantify the inhomogeneous propagation of droplets, we define the spray coefficient ($\gamma$). By employing the ratio of the volumes, i.e., $V_{rc}/V_c$, and $\gamma$,  a scaling factor ($\eta$) can be introduced to obtain the mass of  droplets in $V_{rc}$ as given by 
	\begin{equation}
	\eta = \frac{V_{rc}}{V_c} \gamma = \frac{\frac{\pi}{3} s \left(s \tan{\theta_{rv}}\right)^2}{\frac{\pi}{3} s (s \tan{\theta})^2} \gamma = \left(\frac{\tan{\theta_{rv}}}{\tan{\theta}} \right)^2 \gamma,
	\label{eta}
	\end{equation} 
	where $1 \! \!  \leq \! \!  \gamma \! \!  \leq \! \!  V_c/V_{rc}$. Here, $\gamma \! \!  = \! \!  1$ means that the spatial distribution of droplets are homogeneous within the beamwidth of the outer cone ($2\theta$), and $\gamma \! \! = \! \! V_c/V_{rc}$ means that all droplets propagate in the beamwidth of the inner cone ($2\theta_{rv}$). Then, the droplet concentration in the beamwidth of the RV $(C_0)$ assumed as the initial value of the time-dependent concentration in the RV ($C(t)$), is derived by employing (\ref{eta}) as follows:
	\begin{equation}
	C_0 = \frac{m_{TX} \eta}{V_c} = \frac{m_{TX} \left(\frac{\tan{\theta_{rv}}}{\tan{\theta}} \right)^2 \gamma}{\frac{\pi}{3}s(\frac{R_{rv}}{2})^2} = \frac{3 m_{TX} \gamma }{\pi s^3 \left(\tan{\theta}\right)^2}.
	\label{C_00}
	\end{equation}
	In (\ref{C_00}), $ m_{TX} $ can be presented by using the volumetric flow rate ($ Q $) of the TX which gives the fluid volume flowing through the sprayer per unit time \cite{munson2009fundamentals}. Here, $Q = V_{TX}/T_e$  where $V_{TX}$ shows the emitted liquid volume. Since the transmitted mass  can be written as $m_{TX} = V_{TX}\rho_d$ where $\rho_d$ is the density of the liquid forming droplets before spraying, (\ref{C_00}) is given by
	\begin{equation}
	C_0 = \frac{3 Q T_e \rho_d \gamma }{\pi s^3 \left(\tan{\theta}\right)^2}.
	\label{C_0}
	\end{equation}
	\subsubsection{Signal Reconstruction of the Receiver} \label{SR_RX}
\begin{figure}[h]
		\centering
		\scalebox{0.55}{\includegraphics{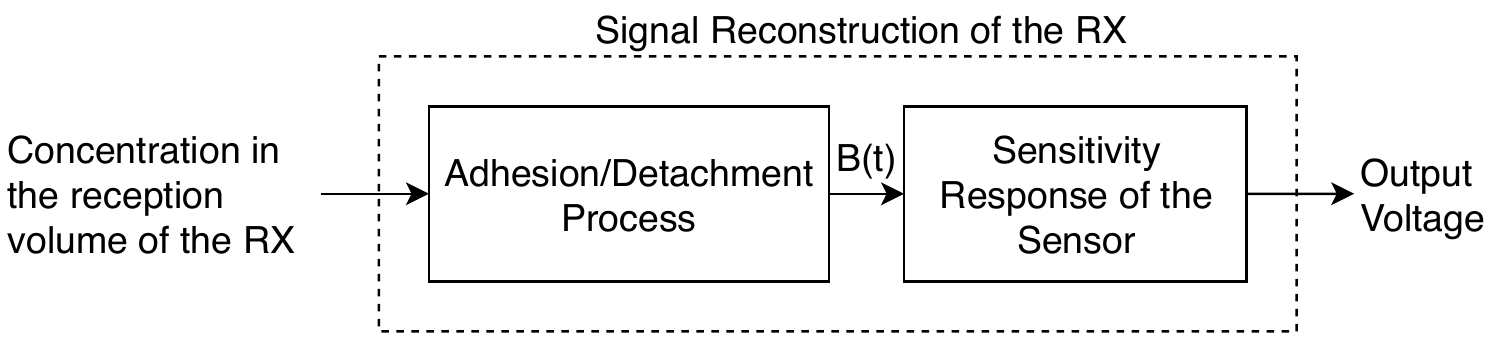}}
		\caption{Block Diagram of the Signal Reconstruction.}
		\label{SR}
	\end{figure}

	For our scenario, the reconstruction of the molecular signal around the RX is subject to an error due to the random adhesions/detachments of droplets to/from the sensor and the sensitivity of the sensor. Hence, the SR is modeled as the combination of the adhesion/detachment process and the sensitivity response of the sensor as shown in Fig. \ref{SR}.
	
	Let $X$ and $Y$ represent the droplets in the RV and the adhered droplet-sensor complex, respectively. Moreover, $Z$ is defined as the detached state of the droplet which is assumed not to be sensed by the sensor. Here, droplets are assumed to arrive the RV instantaneously after the emission in order to relate the chemical kinetics of droplets in the RV and the sensor measurement as a function of time. The adhesion and detachment can be modeled as first order reactions which are given by  
	\begin{equation}
	\ce{X ->[k_1] Y ->[k_2] Z},
	\label{reactions}
	\end{equation}
	where $k_1$ and $k_2$ are the rate constants of the corresponding reaction. Let $ C(t) $ and $ B(t) $ denote the concentrations of $ X $ and $Y$ in kg/m$^3$, respectively. Based on the reaction system in (\ref{reactions}) and the rate law \cite{atkins2010physical},  the concentrations $ C(t) $ and $ B(t) $ can be characterized as 
	\begin{align}
	\frac{d C(t)}{dt} &= -k_1 C(t) \label{reaction1}\\
	\frac{d B(t)}{dt} &= k_1 C(t) - k_2 B(t),\label{reaction2}
	\end{align}
	where the initial conditions are defined as $C(0) = C_0$ and $B(0) = 0$. 
	The solution  of (\ref{reaction1})-(\ref{reaction2}) for $B(t)$ can be given as 
	\begin{equation}
	B(t) = \frac{k_1 C_0}{k_2-k_1} [e^{-k_1t}-e^{-k_2t}].
	\label{B}
	\end{equation}
	
\begin{figure}[h]
		\centering
		\scalebox{0.65}{\includegraphics{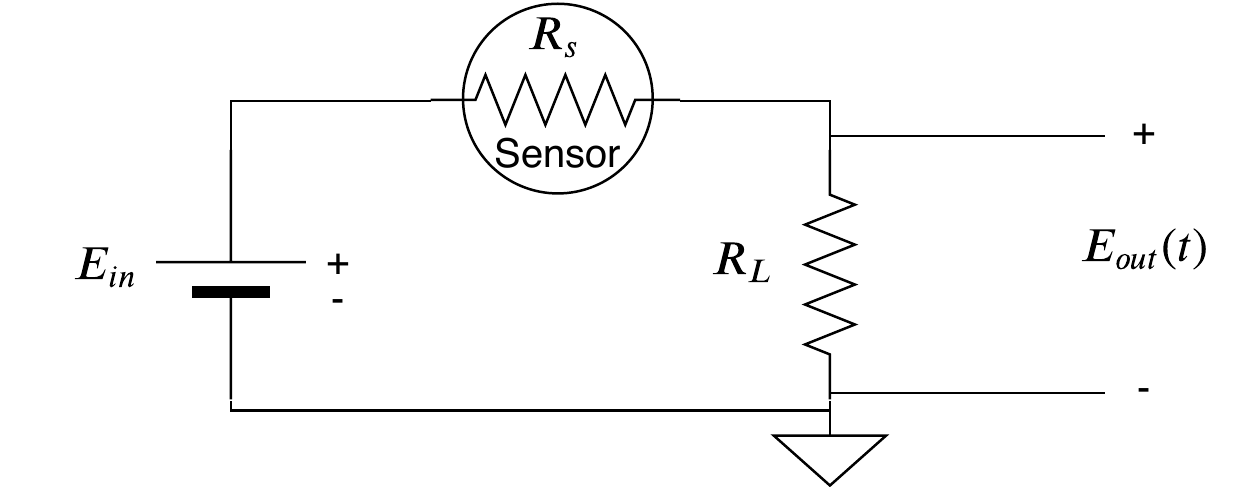}}
		\caption{Sensor Measurement Circuit.}
		\label{Circuit}
	\end{figure}

	Subsequent to adhesion/detachment process, $ B(t) $ is converted to an electrical signal by the metal-oxide MQ-3 sensor. Sensors of this type measure the concentration around them by changing their resistance so that each concentration value corresponds to a sensor resistance ($R_S$) value as shown in Fig. \ref{Circuit}. In order to normalize the measured $R_s$, $R_o$ is defined as the sensor resistance measured at $ 0.0004$ kg/m$^3$ \cite{MQ3}. For each concentration value, $R_S/R_o$ determines how sensitive the sensor can measure. This sensitivity characteristic, whose values are taken from its datasheet \cite{MQ3}, can be employed to obtain a sensitivity function $\left(f(t)\right)$ which maps concentration values to $R_S/R_o$ values via curve fitting technique. The datasheet values are fitted by using nonlinear least squares method which minimizes the sum of square errors. $f(t)$ can be fitted as given by
	\begin{equation}
	f\left(B\left(t\right)\right) =  0.0116 \left(B(t)\right)^{-0.5855} - 0.0743 = \frac{R_S}{R_o},
	\label{f}
	\end{equation}
	where $B(t)$ is the input to the sensitivity response of the sensor as shown in Fig. \ref{SR} and derived in (\ref{B}). The scalar curve fitting  parameters are estimated by employing  Levenberg-Marquardt algorithm  which has an estimation performance with the Root Mean Square Error (RMSE) value of $ 0.0371 $ \cite{hagan1994training}. 
	
	In our experimental setup, the sensor measurement is made with a circuit as given in Fig. \ref{Circuit} where $R_L$ is the load resistance, $E_{in}$ is the input voltage and $E_{out}(t)$ is the output voltage which also gives the end-to-end system impulse response for an impulsive input signal. Using this circuit, $R_S$ can be derived via Kirchhoff's voltage law \cite{nilsson2010electric} as given by
	\begin{equation}
	R_S = \left(\frac{E_{in}}{E_{out}(t)} - 1\right) R_L,
	\label{R_S}
	\end{equation}
	where $E_{in}$ is given as $5$ V in \cite{MQ3}. By combining  (\ref{f}) and (\ref{R_S}), the relation between $f(B(t))$ and the parameters of the sensor measurement circuit  can be written as
	\begin{equation}
	f(B(t)) =  \left(\frac{E_{in}}{E_{out}(t)} - 1\right) \frac{R_L}{R_o}.
	\label{f2}
	\end{equation}
	By using (\ref{f2}), $ E_{out}(t) $ is given as 
	\begin{equation}
	E_{out}(t) =  \dfrac{E_{in}R_L}{R_o\left(f(B(t))+\frac{R_L}{R_o}\right)}.
	\label{E_out}
	\end{equation}
	Finally, the end-to-end impulse response can be expressed by substituting (\ref{B}) into (\ref{E_out}) as given by
	\begin{equation}
	E_{out}(t) = \dfrac{E_{in}R_L}{R_o\left[f\left( \frac{k_1 C_0}{k_2-k_1} ( e^{-k_1 t}-e^{-k_2 t} )       \right) + \frac{R_L}{R_o}\right]},
	\label{h}
	\end{equation}
	where it is important to note that $C_0$ is a function of the spray coefficient $\gamma$ as given in (\ref{C_0}). Therefore, the impulse response in (\ref{h}) involves three novel channel parameters, i.e., $k_1$, $k_2$ and $\gamma$. These parameters are affected by the change of other parameters in the channel such as the distance between the TX and RX. Especially, $k_1$ is a function of the average velocity, which is defined as the average velocity of droplets entering the RV, since the flow stemming from the sprayer causes droplets to react with the sensor surface.  Besides, $\gamma$ depends on the type of the sprayer's nozzle which affects the spraying pattern \cite{al2014influence} and interactions among droplets and air molecules. For numerical results, $k_1$, $k_2$, and $\gamma$ are manually configured by making the Mean Square Error ($\epsilon$) between the samples of $E_{out}(t)$ and experimental signal ($F(t)$) as small as possible according to the formula which is given as
	\begin{equation}
	\epsilon = \frac{1}{N}\sum_{n=1}^{N}\left(E_{out}[n] - F[n] \right)^2
	\end{equation}
	where $N$ shows the number of samples, $E_{out}[n]$ and $F[n]$ are discrete-time representations of $E_{out}(t)$ and $F(t)$, respectively. Next, it is shown by numerical results that the proposed model given in (\ref{h}) can be used for practical scenarios.
	\section{Numerical Results}\label{Exp_3}
	\subsection{Measurements}
	In order to measure $ Q $, the sprayer is placed on a precision balance. For each measurement, the liquid is sprayed from the sprayer for a short interval ($\Delta t_v$). The mass of the sprayer is measured before and after spraying. Hence, the mass difference is found by measuring the mass values before and after spraying. By dividing the mass difference to $ \rho_d $, the volume difference ($ \Delta V $) is calculated for each measurement. Thus, $Q$ can be presented by \cite{munson2009fundamentals}
	\begin{equation}
	Q = \frac{\Delta V}{\Delta t_v}.
	\label{Q}
	\end{equation}
	Here, the average of ten measurements is considered for $Q$. In addition, $R_o$ is calculated by using $E_{out}(t)$ value when the concentration value is  $ 0.0004$  kg/m$^3$. The detection scope of the sensor which is between $5\times10^{-5}$ and $10^{-2}$ kg/m$^3$, is scaled for $E_{out}(t)$ values between $0$ and $5$ V \cite{MQ3}. Hence, $E_{out}(t) = 0.2$ V for the concentration value of $ 0.0004$ kg/m$^3$. Then, $R_o$ is calculated by (\ref{R_S}). Furthermore, $\theta$ is measured with ImageJ software using the image given in Fig. \ref{RV}. Next, the values of the experimental parameters given in Table \ref{Sim_parameters} are used for the validation of the proposed model.
	\begin{table}[t]
		\centering
		\caption{Experimental parameters}
		\scalebox{0.9}{
			\begin{tabular}{ll}
				\hline
				\textbf{Parameter}	& \textbf{Value} \\ 
				\hline  \hline
				Distance between the TX and RX ($ s $) & $\{0.9, 1, 1.1, 1.2\}$ m \\
				Volumetric flow rate ($ Q $) &  $2.204 \times 10^{-6}$ m$^3$/s  \\
				Density of liquid ethanol ($\rho_d$) & $ 789$ kg/m$^3 $ \\
				Emission time of the TX ($T_e$) &  $ 0.5$ s \\
				Load resistance ($R_L$) & $1$ k$\Omega$ \\
				Sensor resistance at $ 0.0004$ kg/m$^3$ ($ R_o $) & $24$ k$\Omega$ \\
				Half-beamwidth of the sprayer ($\theta$) & $38^\circ$ \\
				\hline   \hline           
		\end{tabular}}
		\label{Sim_parameters}
	\end{table}
	\subsection{Results for Channel Modeling}\label{Numerical_Channel}
\begin{figure*}[b]
		\centering
		\includegraphics[width=0.243\textwidth]{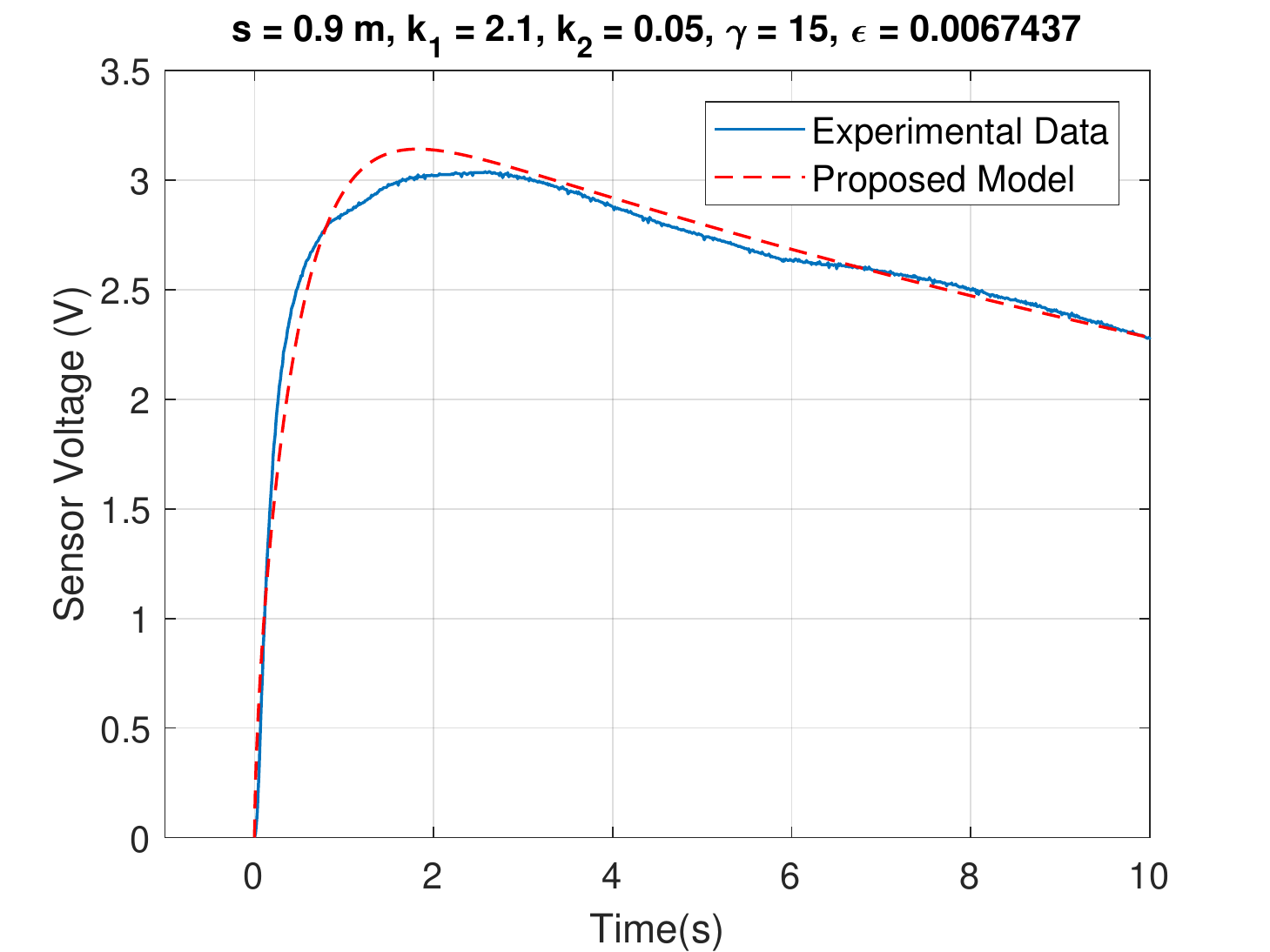}    
		\includegraphics[width=0.243\textwidth]{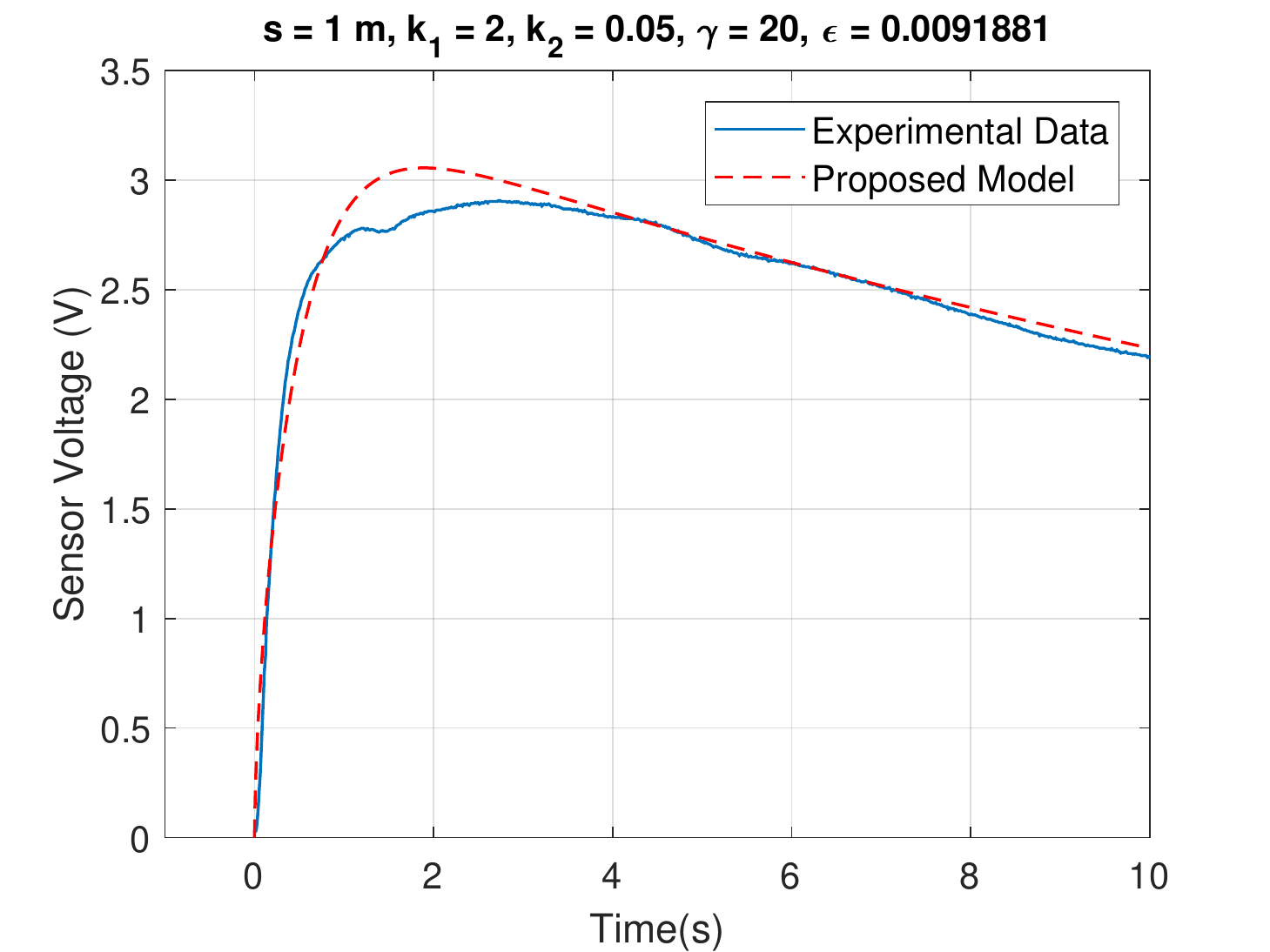}  
		\includegraphics[width=0.243\textwidth]{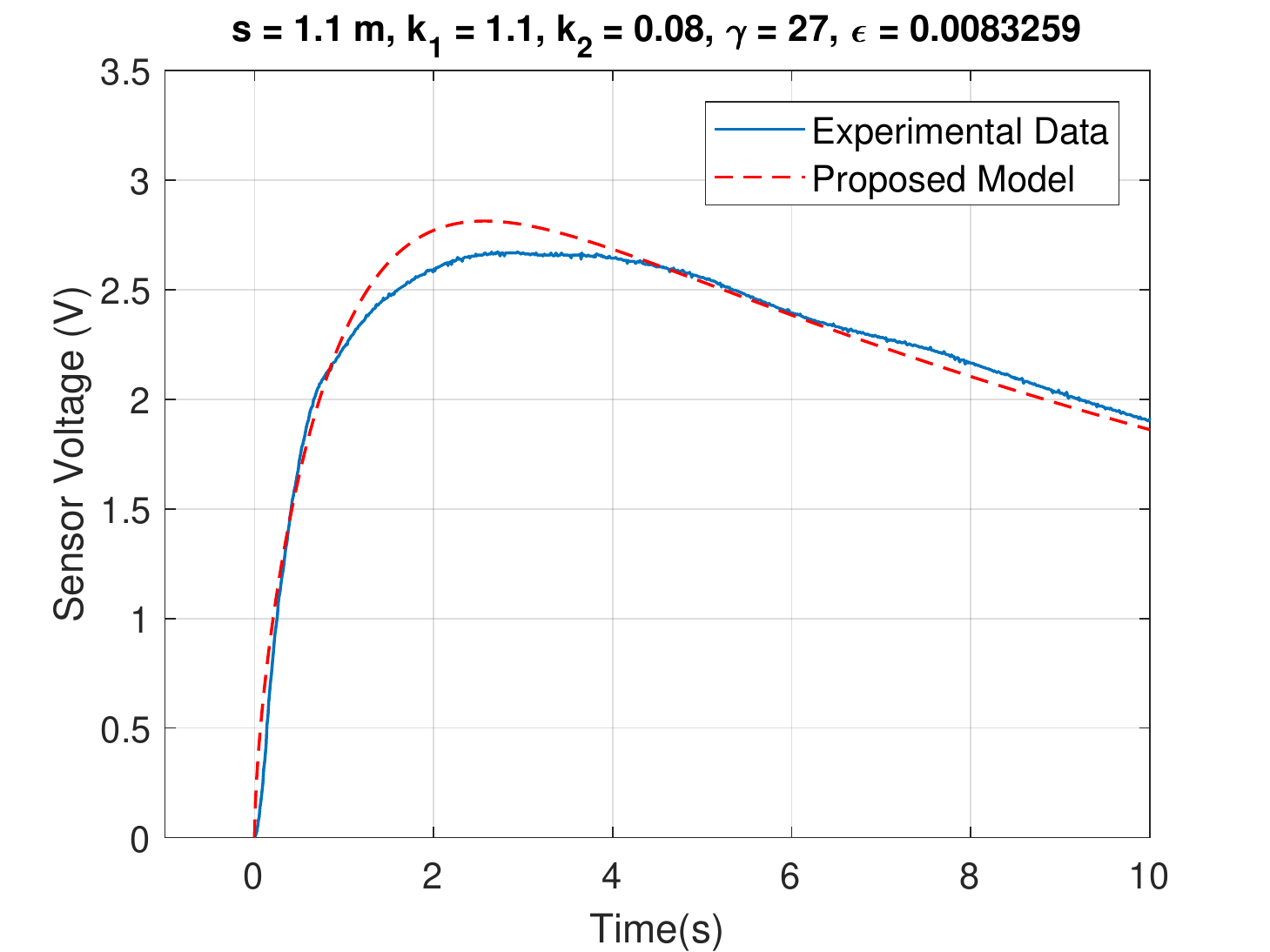} 
		\includegraphics[width=0.243\textwidth]{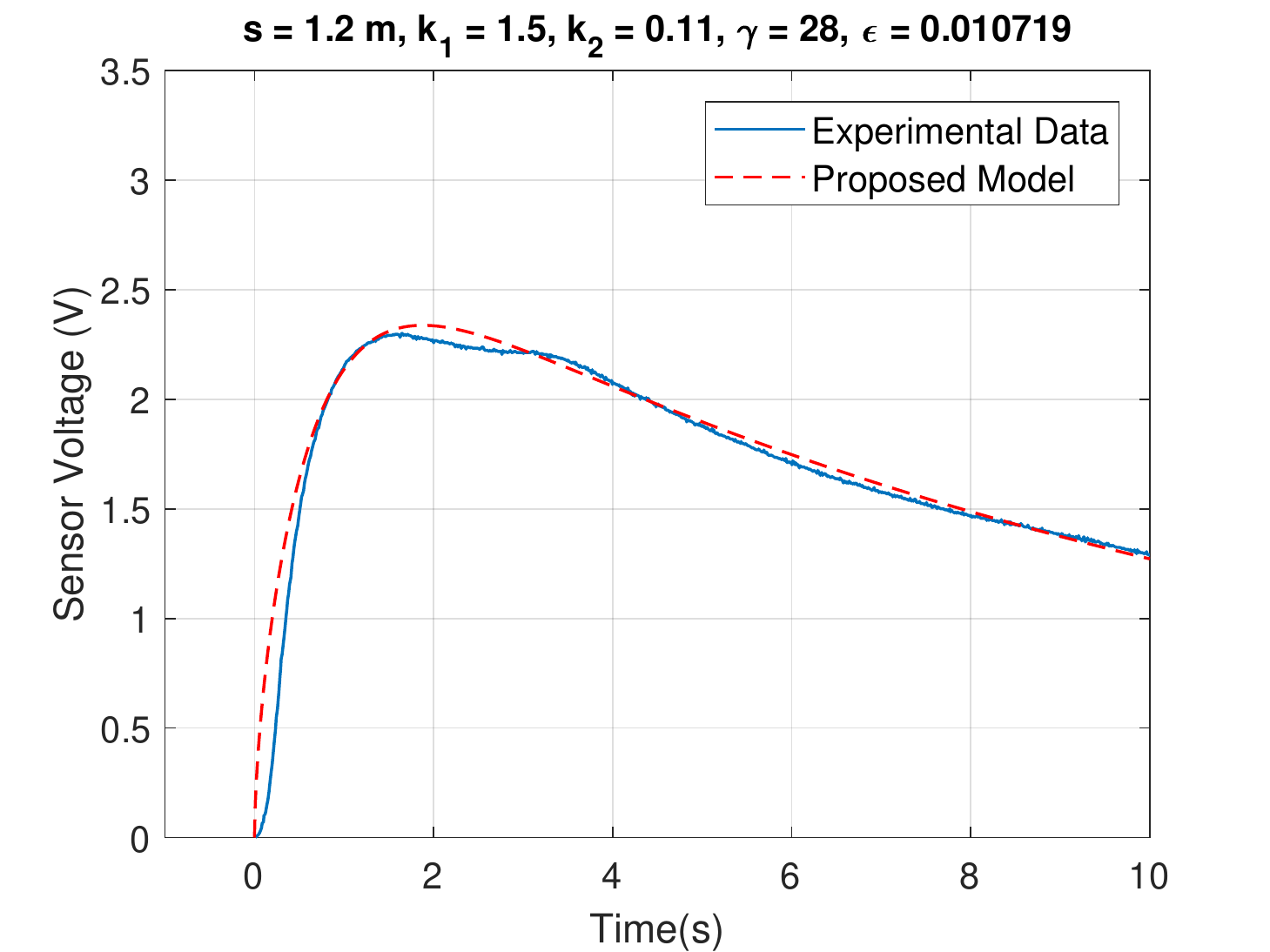} \\
		\scriptsize \hspace{0.14 cm} (a) \hspace{3.63 cm}  (b) \hspace{3.63 cm} (c) \hspace{3.63 cm} (d) \\
		\caption{The comparisons of the proposed model with experimental data for the parameters given in the title of each signal.}
		\label{d_signals}
	\end{figure*}

	In Fig. \ref{d_signals}, the comparisons of experimental data and the proposed model are shown. For performance evaluation, $\epsilon$ is calculated for a duration of $10$ s. Fig. \ref{d_signals} validates that the proposed channel model can be employed to estimate the received signal.
	

	For our macroscale scenario, $k_1$ and $k_2$ represent the average characteristic of the random movements of droplets in the RV. In addition, $\gamma$ describes the effect of the propagation of droplets on the initial concentration in the RV. As shown in Fig. \ref{d_signals}, $ \gamma $ increases as $ s $ increases, since the shape of droplets' spatial dispersion gets narrower with $s$. The increment of $\gamma$  with $s$ validates the narrowing beamwidth at longer distances due to the droplet-air interaction \cite{ghosh1994induced}. Moreover, the slope of the rising edge of the signals in Fig. \ref{d_signals} between $ 0 $ and peak time is proportional to $ k_1 $. This slope decreases as $ s $ increases, since $ k_1 $ is proportional to the average velocity of droplets entering the RV. In addition, as $k_2$ increases, the adhered droplets on the sensor detach faster and the sensor voltage drops faster from its peak value to its initial level. 
	\begin{figure}[t]
		\centering
		\scalebox{0.4}{\includegraphics{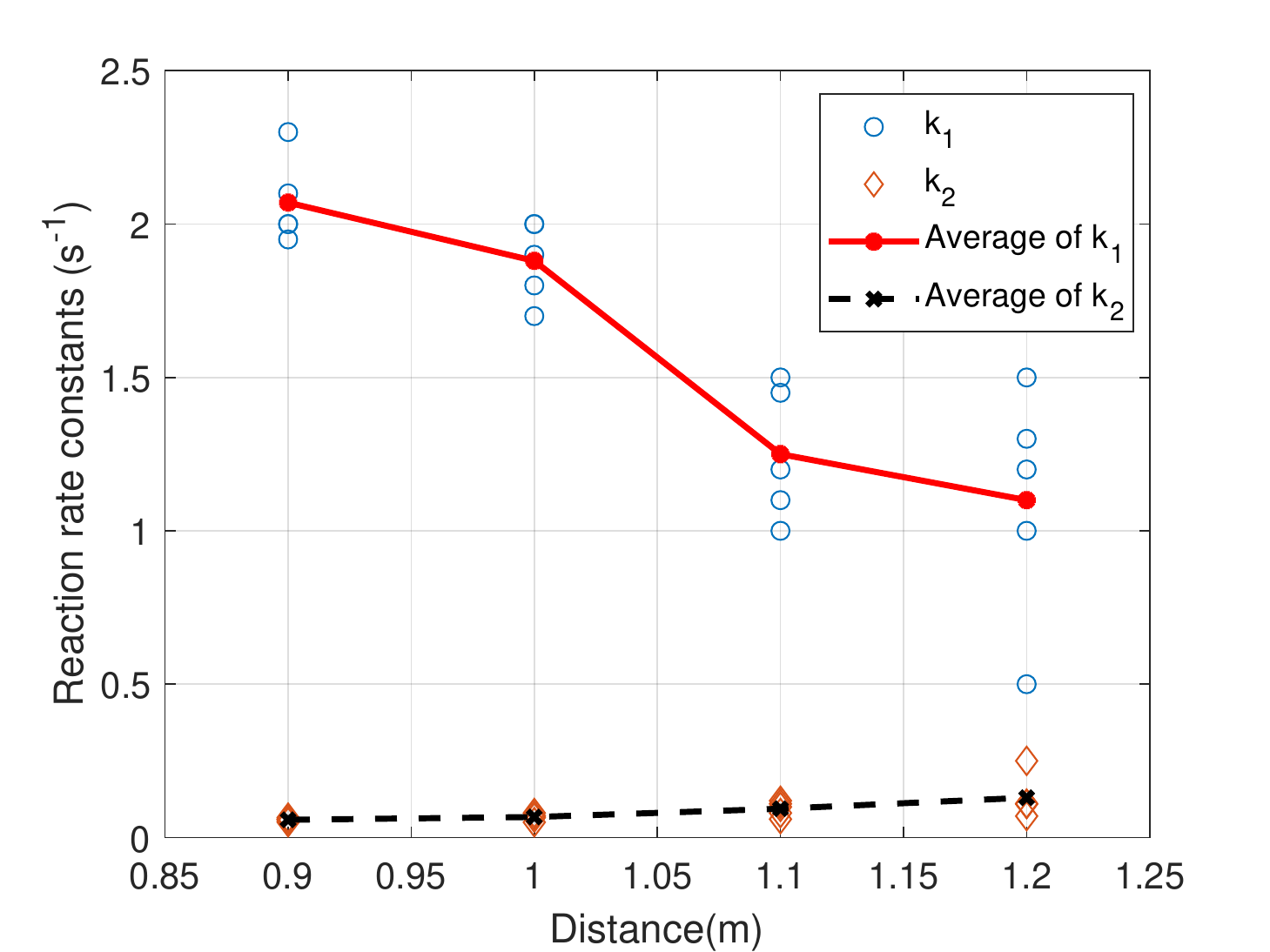}}
		\caption{The relation of $ k_1 $ and $ k_2 $ with the distance.}
		\label{k_12}
	\end{figure} 
	In Fig. \ref{k_12}, twenty different manually fitted signals (with a $\epsilon$ of less than $ 0.021 $) are used to observe the relation of $k_1$ and $k_2$ with the distance. Droplets are more unlikely to stay adhered to the sensor due to the decreasing average velocity for longer distances, as shown with the change of average $k_1$ values in Fig. \ref{k_12}. In contrast to $k_1$, $k_2$ is almost constant, since the average velocity of droplets is not so effective on their detachment from the sensor in the RV.
	\section{Conclusion}
	\label{Conclusion}
	In this paper, an end-to-end system model is proposed for macroscale MC systems with a fluid dynamics approach. This approach is merged with the SR of the RX which considers the adhesion/detachment of droplets and the sensor's sensitivity. This study reveals the physical meanings  of the channel parameters. As the future work, we plan to employ fluid dynamics approach for channel parameter estimation.

	\bibliographystyle{ieeetran}
	\bibliography{ref_fg}

\end{document}